\documentclass[twocolumn]{aastex631}
\usepackage{txfonts}
\usepackage{newtxtext}
\usepackage{mathrsfs}
\usepackage[T1]{fontenc}
\usepackage{xcolor}
\usepackage{color}
\usepackage{soul}
\usepackage{float}
\usepackage{hyperref}
\usepackage[graphicx]{realboxes}
\usepackage[toc,page]{appendix}

\begin{document}

\title{COSMIC-S: a photometric Catalogue of Observed Stars in the Small MagellanIc Cloud}

\correspondingauthor{Antonio Franco}
\email{antonio.franco@le.infn.it}

\author[0000-0002-4761-366X]{Antonio Franco}
\affiliation{INFN, Sezione di Lecce, Via per Arnesano, CP-193, I-73100, Lecce, Italy}
\affiliation{INAF, Sezione di Lecce, Via per Arnesano, CP-193, I-73100, Lecce, Italy}
\affiliation{Department of Mathematics and Physics {\it ``E. De Giorgi''} , University of Salento, \\ Via per Arnesano, CP-I93, I-73100, Lecce, Italy}

\author[0000-0002-7926-3481]{Achille A. Nucita}
\affiliation{INFN, Sezione di Lecce, Via per Arnesano, CP-193, I-73100, Lecce, Italy}
\affiliation{INAF, Sezione di Lecce, Via per Arnesano, CP-193, I-73100, Lecce, Italy}
\affiliation{Department of Mathematics and Physics {\it ``E. De Giorgi''} , University of Salento, \\ Via per Arnesano, CP-I93, I-73100, Lecce, Italy}

\author[0000-0001-6460-7563]{Francesco De~Paolis}
\affiliation{INFN, Sezione di Lecce, Via per Arnesano, CP-193, I-73100, Lecce, Italy}
\affiliation{INAF, Sezione di Lecce, Via per Arnesano, CP-193, I-73100, Lecce, Italy}
\affiliation{Department of Mathematics and Physics {\it ``E. De Giorgi''} , University of Salento, \\ Via per Arnesano, CP-I93, I-73100, Lecce, Italy}

\author[0000-0002-8757-9371]{Francesco Strafella}
\affiliation{INFN, Sezione di Lecce, Via per Arnesano, CP-193, I-73100, Lecce, Italy}
\affiliation{INAF, Sezione di Lecce, Via per Arnesano, CP-193, I-73100, Lecce, Italy}
\affiliation{Department of Mathematics and Physics {\it ``E. De Giorgi''} , University of Salento, \\ Via per Arnesano, CP-I93, I-73100, Lecce, Italy}

\begin{abstract}

The Dark Energy Camera (DECam) is a wide-field instrument mounted on the 4m V. Blanco Telescope (CTIO). Its impressive technical characteristics makes it one of the most suitable ground-based telescope for the production of accurate stellar photometry even towards crowded regions such as the Magellanic Clouds.
    
We analysed DECam images acquired from February 2018 to January 2020 towards the Small Magellanic Cloud. We performed a PSF photometry by using the SExtractor and PSFEx tools and producing a comprehensive photometric catalogue in the SDSS system, considering the $gri$ filters.

Then, we present COSMIC-S, a photometric catalogue consisting of 10~971~906 sources, including $gri$ magnitudes with a mean error $<\sigma_m> \simeq 0.04$~mag. A total of 2~456~434 sources have good photometry in all three bands. The catalogue appears virtually complete to $m \simeq 22$, with a limiting magnitude $m\simeq 25$. We derived the colour-magnitude and colour-colour diagrams in order to prove the goodness of the catalogue.

\end{abstract}

   

\section{Introduction}
The Magellanic Clouds (MCs), two irregular dwarf galaxies, satellites of our own, are extremely interesting regions where stellar birth, evolution and death can be studied in detail together with stellar variability. The MCs have been the main focus of many astronomical surveys, such as the Dark Energy Survey (DES, \citealt{sanchez2010}), started in 2013 with the main purpose of estimating the amount of dark energy responsible for the accelerated expansion of the Universe. In particular, the Small Magellanic Cloud (SMC in the following) offers the possibility to study a galaxy having a chemical composition poorer in metals as compared to the Large Magellanic Cloud (LMC). Moreover, as previously mentioned, the SMC, together with the LMC, represents one of the best scenarios in which the study of variability, e.g. including variable stars (see, e.g., \citealt{franco2023, soszynki2015a, soszynki2015b, soszynki2016, soszynki2017, soszynki2018, udalski2015}), transients phenomena (e.g. microlensing, see \citealt{calchinovati2013, franco2024, mroz2024a, mroz2024b}) and young stellar objects, can be effectively conducted thanks to the presence of numerous suitable regions with diverse evolutionary stages.

\begin{figure*}[ht!]
    \centering
    \includegraphics[width=0.3\textwidth]{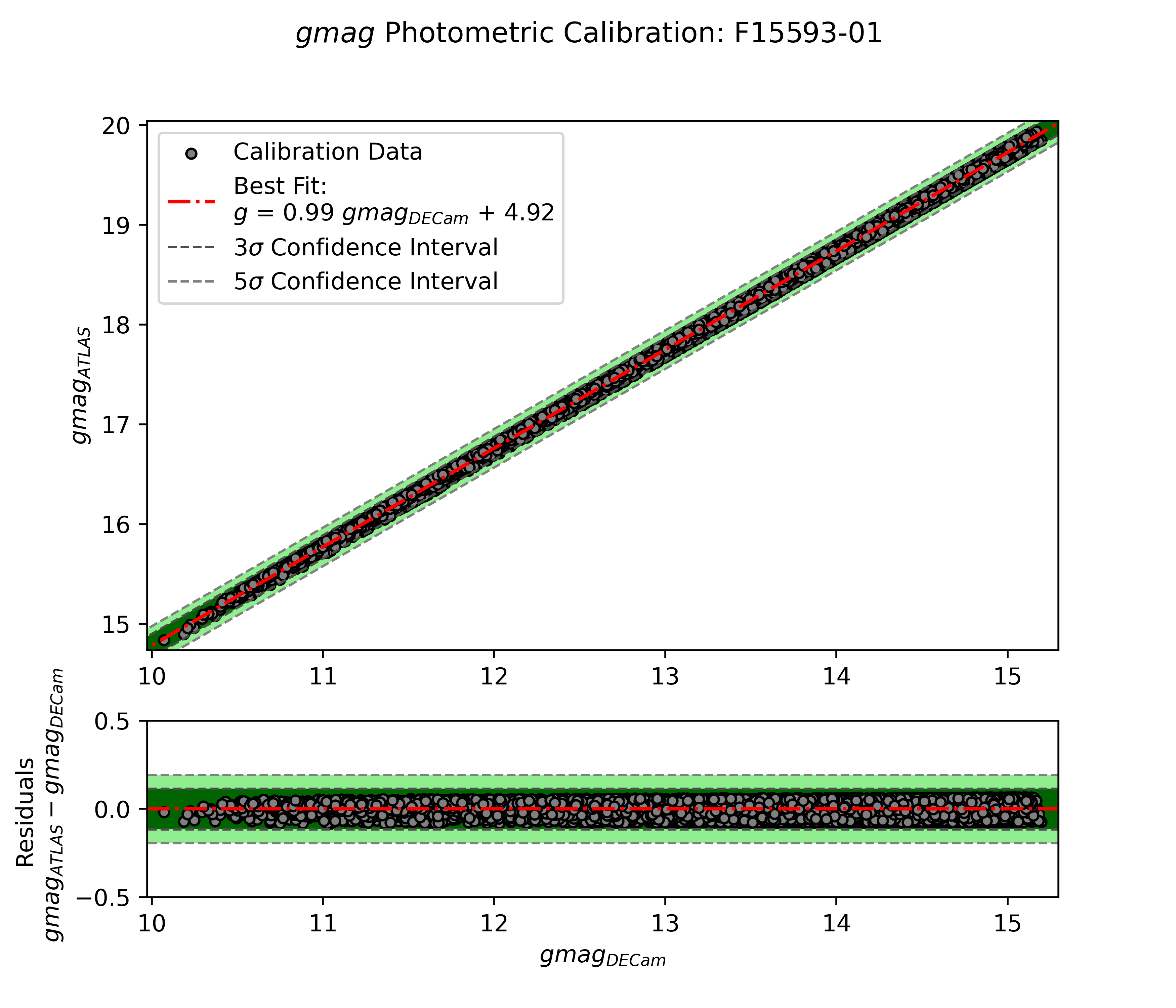}
    \includegraphics[width=0.3\textwidth]{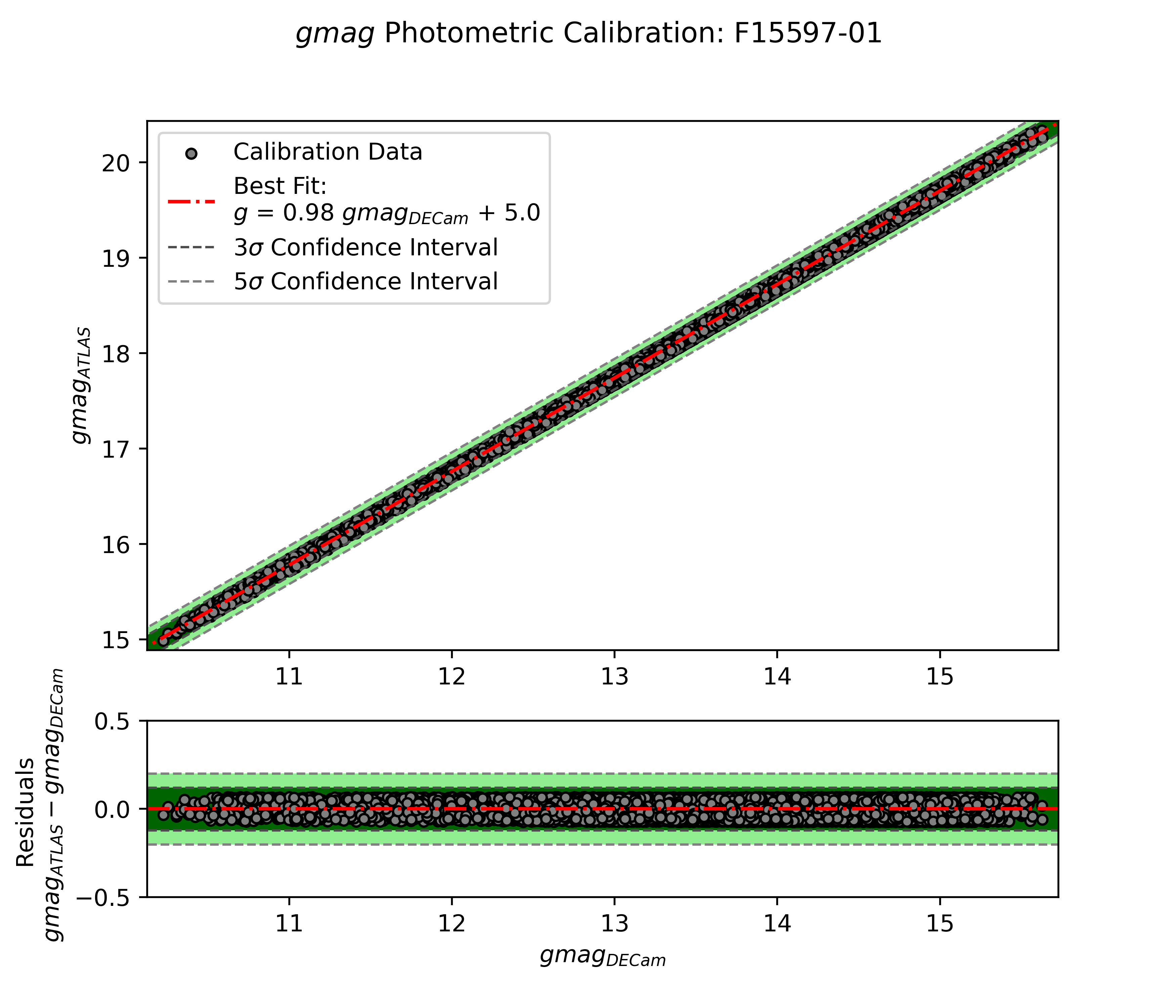}
    \includegraphics[width=0.3\textwidth]{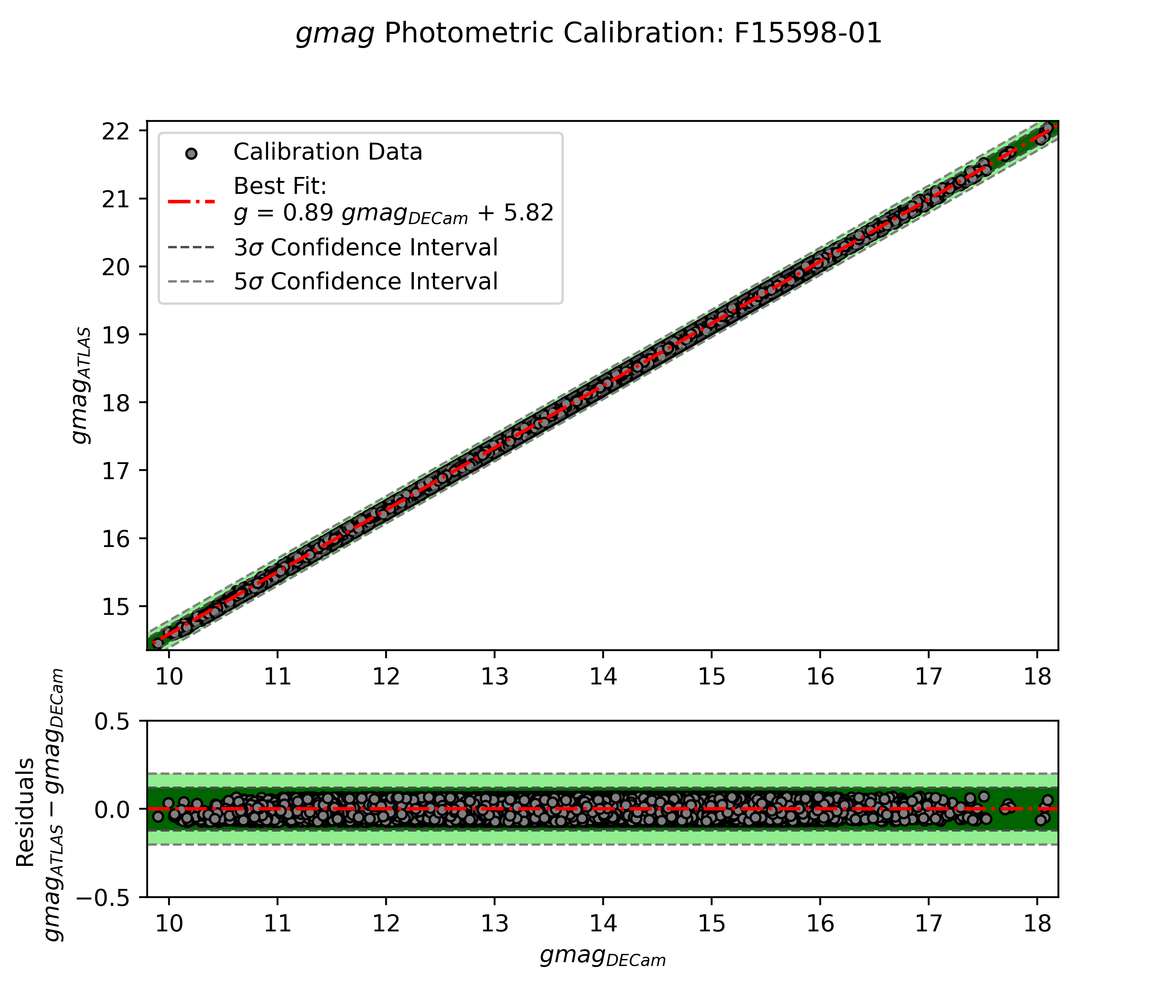}
    \includegraphics[width=0.3\textwidth]{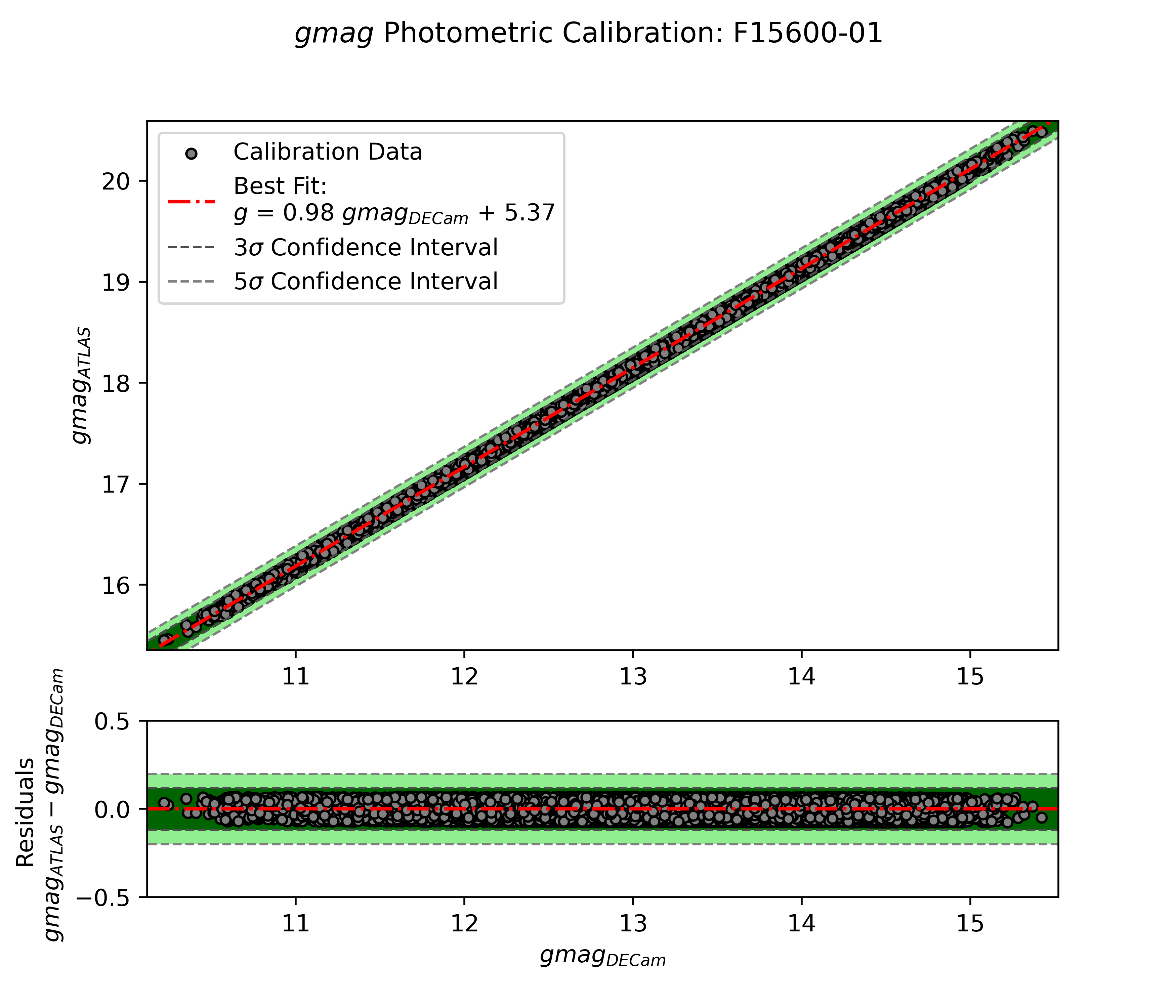}
    \includegraphics[width=0.3\textwidth]{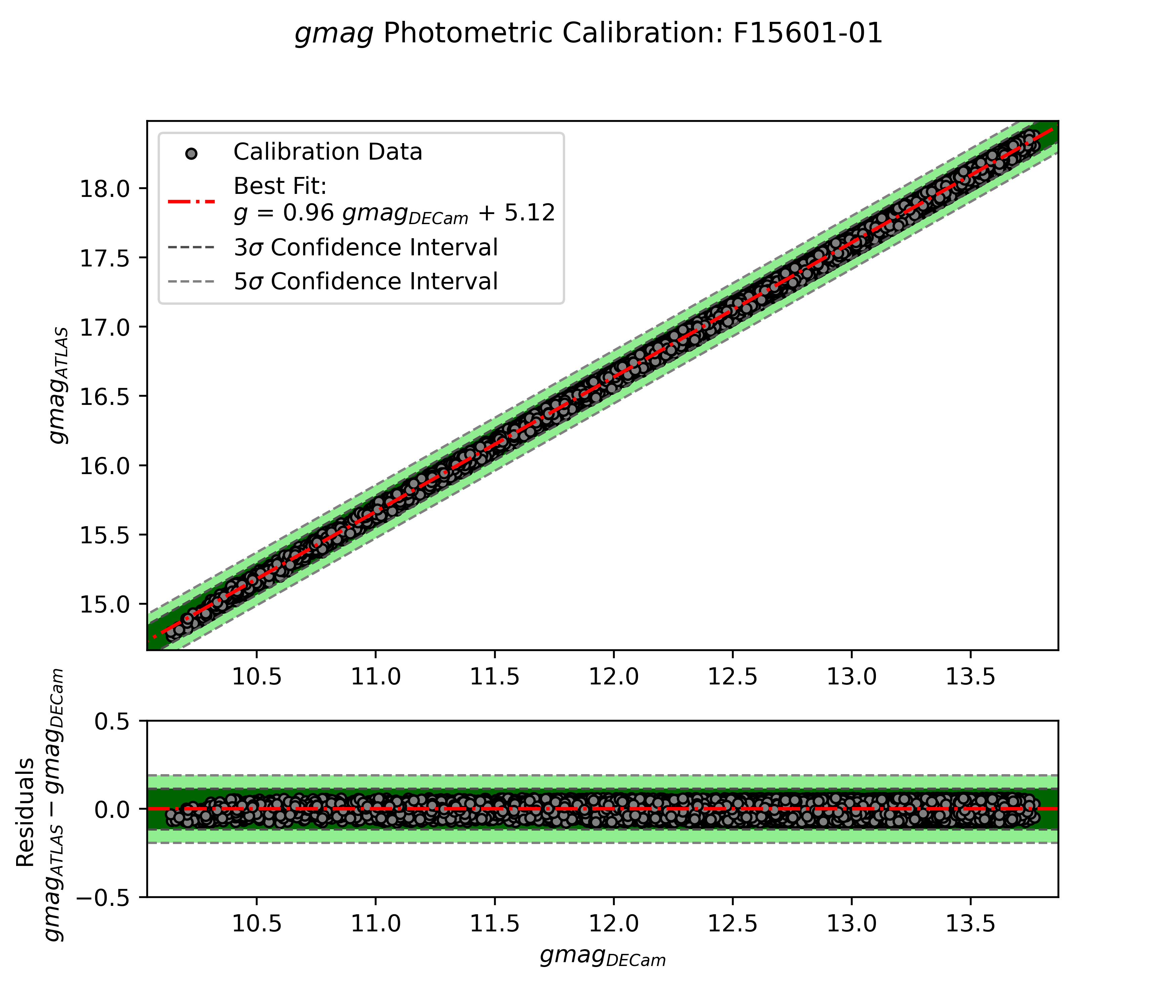}
    \includegraphics[width=0.3\textwidth]{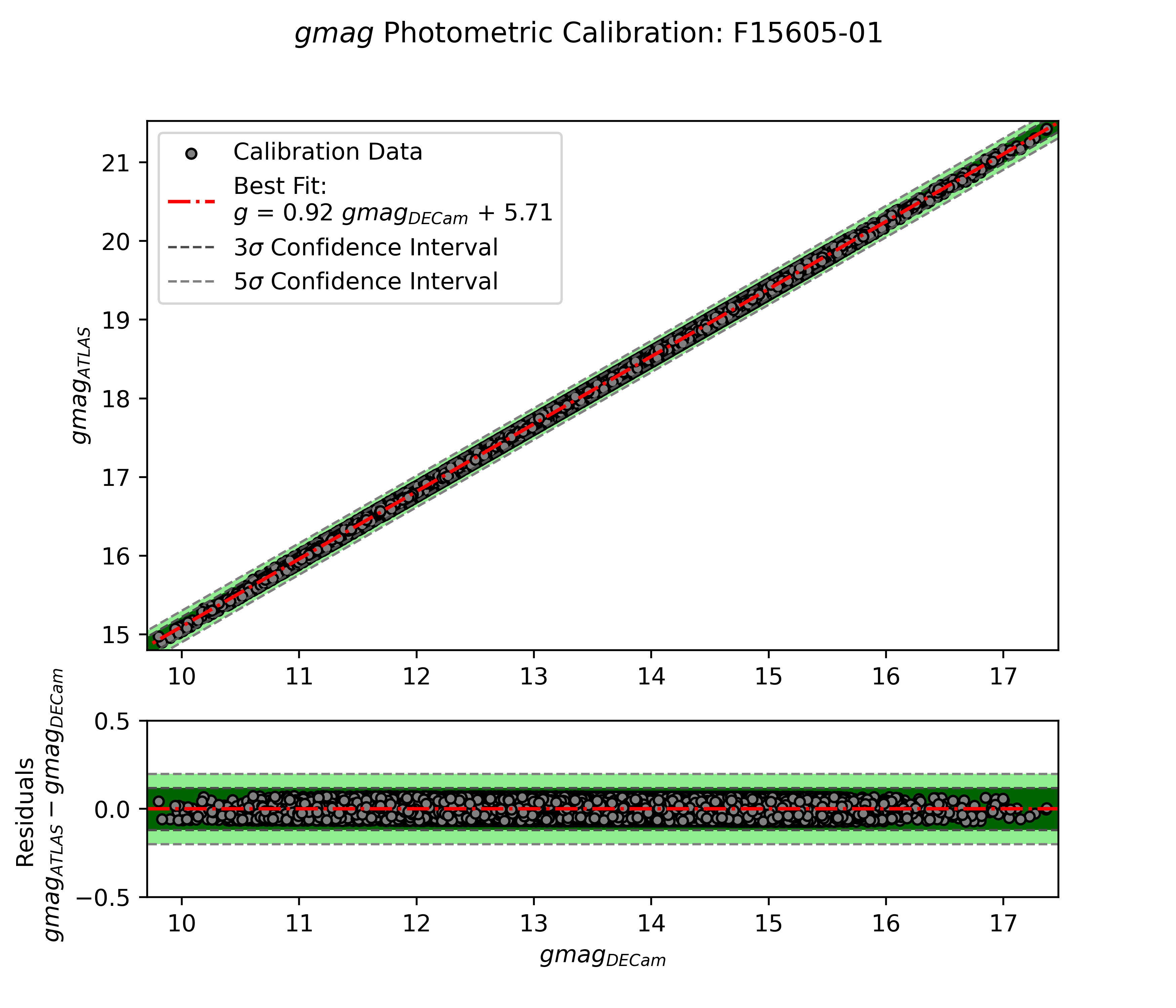}
    \caption{Photometric calibration of the $g$-band images for all the six DECam fields towards the SMC. Each panel refers to the calibration of one DECam field. For each panel, the upper box shows the DECam magnitudes on the x-axis and the ATLAS magnitudes, used for the comparison, on the y-axis. Data used for the calibration are plotted as grey dots. The red dash-dotted line indicates the resulting best fit, as also indicated in the upper left legend panel. The two shades of green indicate the $3\sigma$ (darker) and $5\sigma$ (lighter) confidence interval. The bottom box shows instead the fit residuals. The shades of green have the same previous meaning. The red dash-dotted line represents corresponds to the zero residual line.}
    \label{Fig::calib_g}%
\end{figure*}

\begin{figure*}[ht!]
    \centering
    \includegraphics[width=0.3\textwidth]{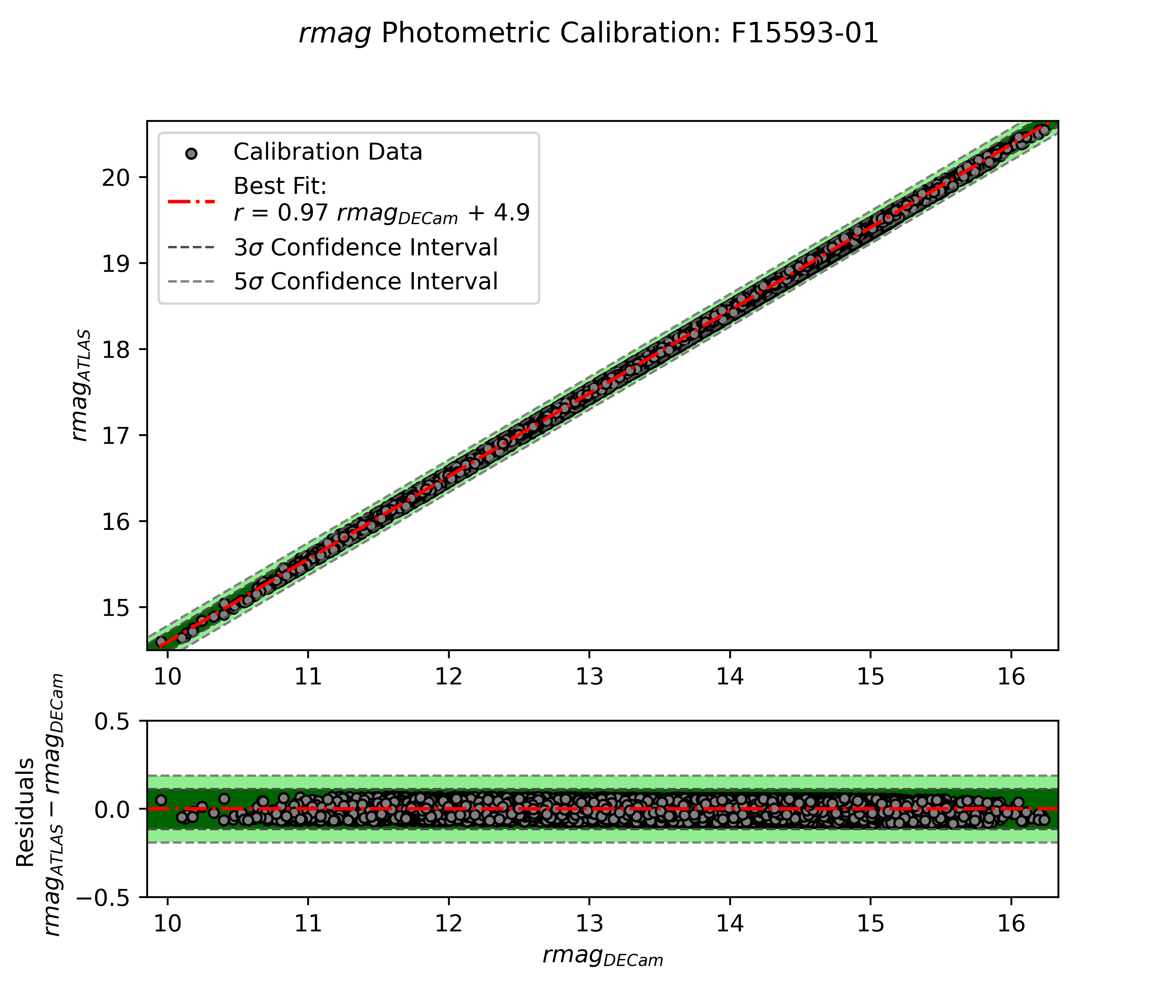}
    \includegraphics[width=0.3\textwidth]{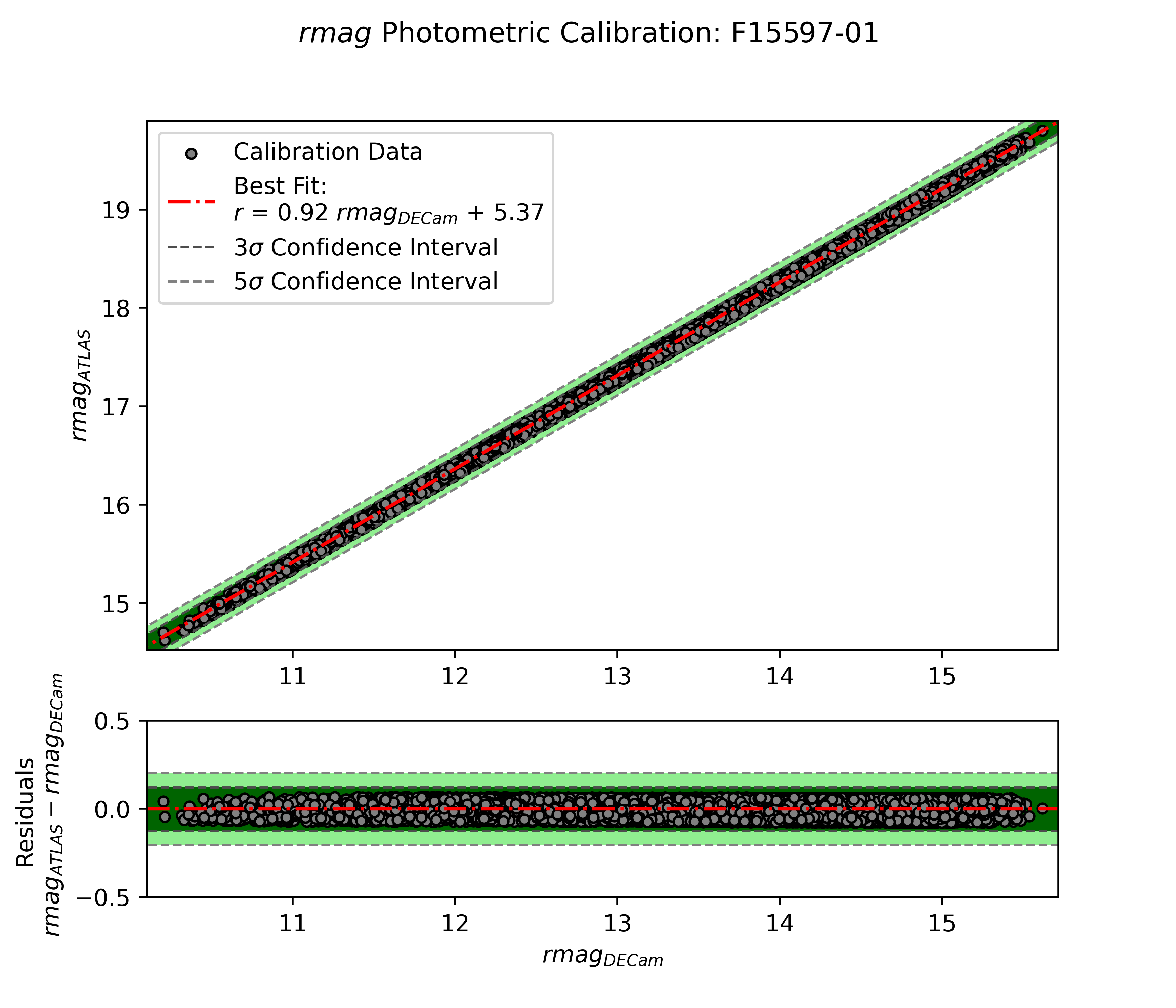}
    \includegraphics[width=0.3\textwidth]{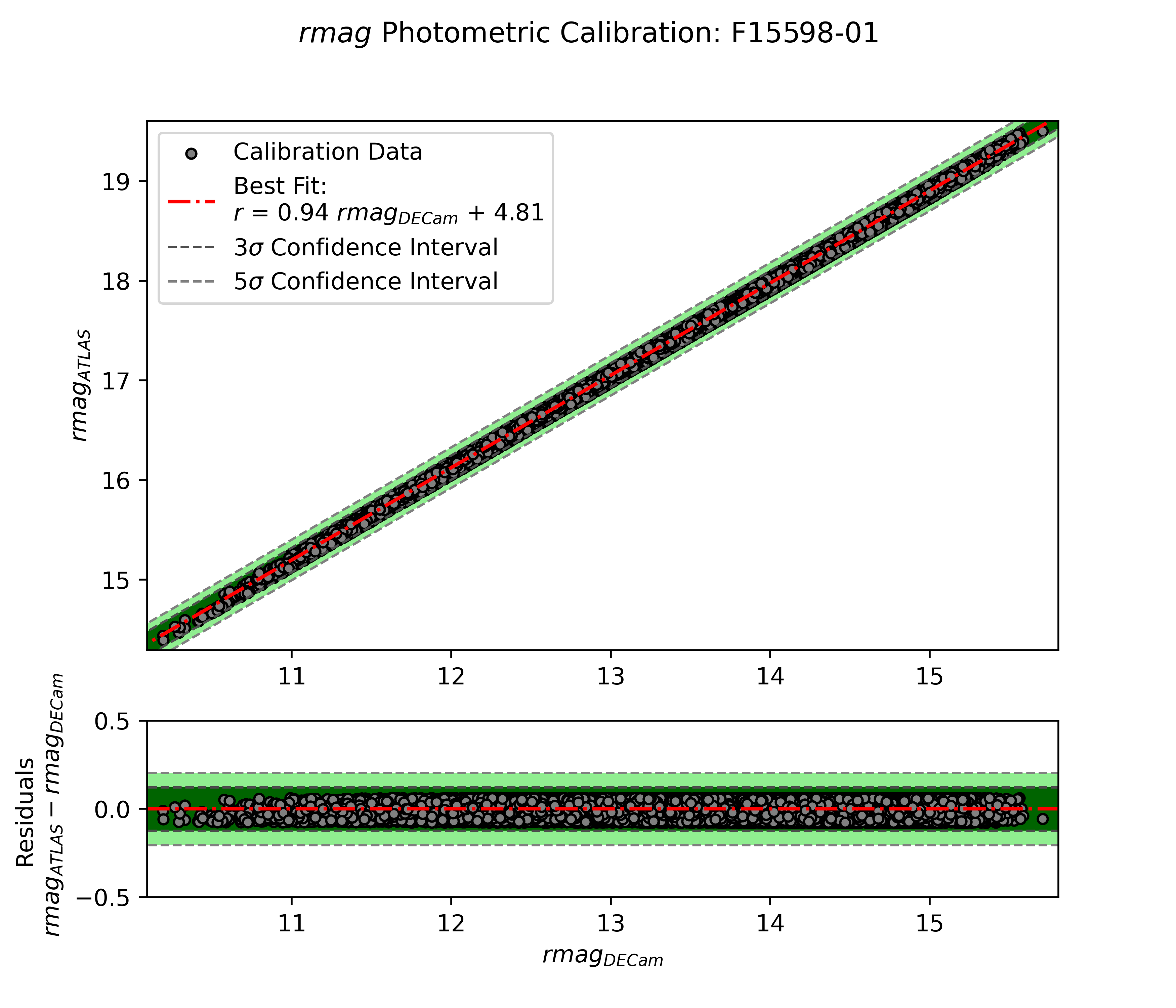}
    \includegraphics[width=0.3\textwidth]{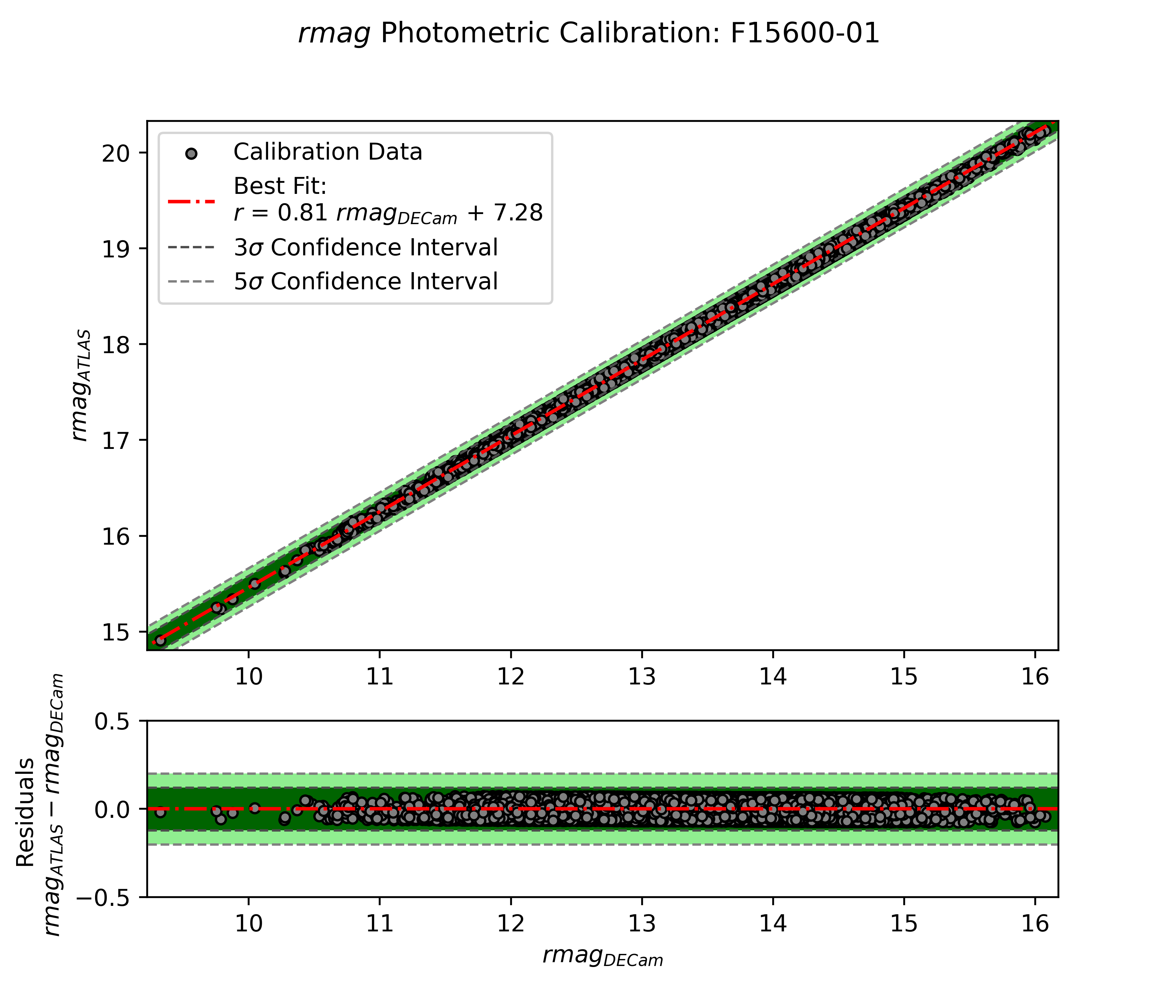}
    \includegraphics[width=0.3\textwidth]{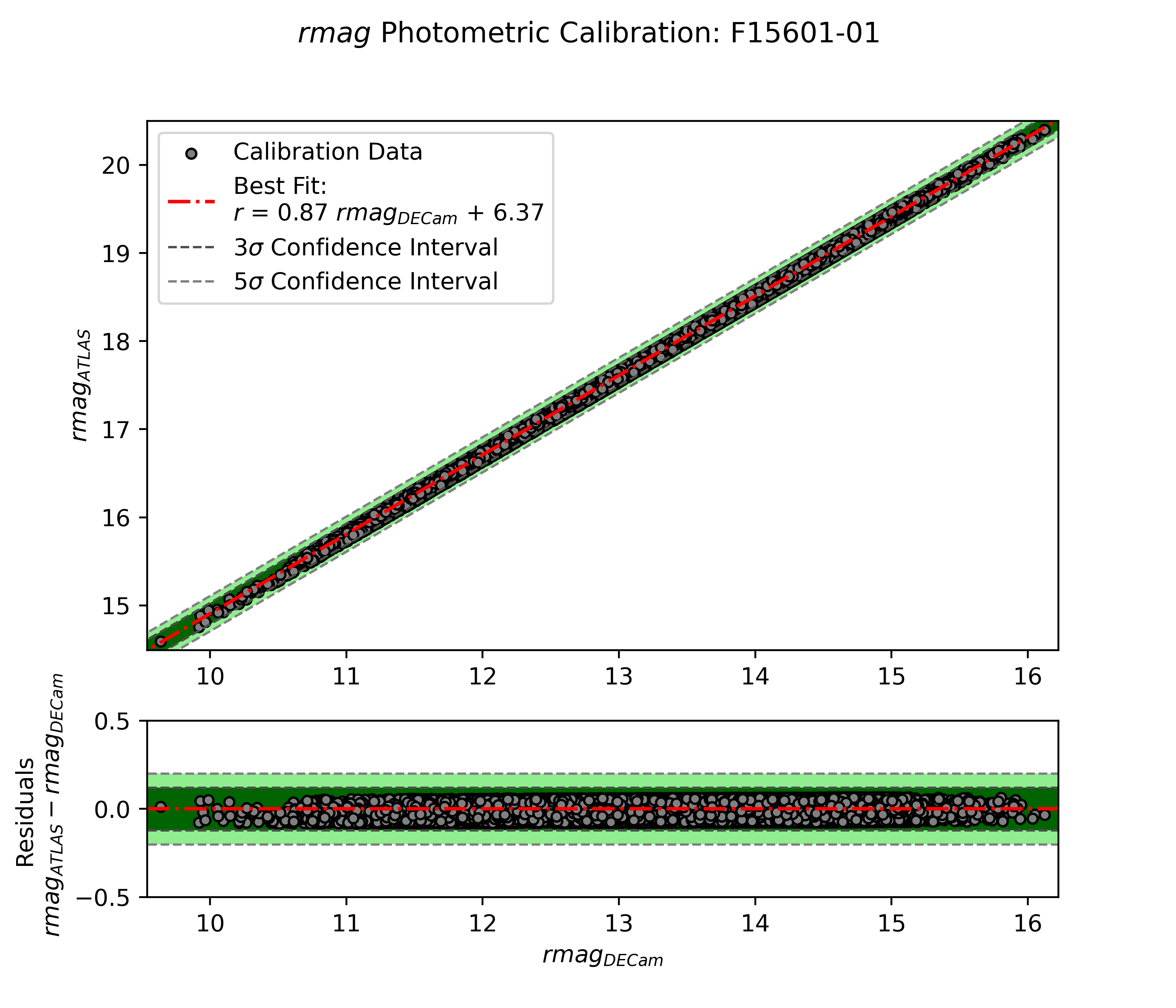}
    \includegraphics[width=0.3\textwidth]{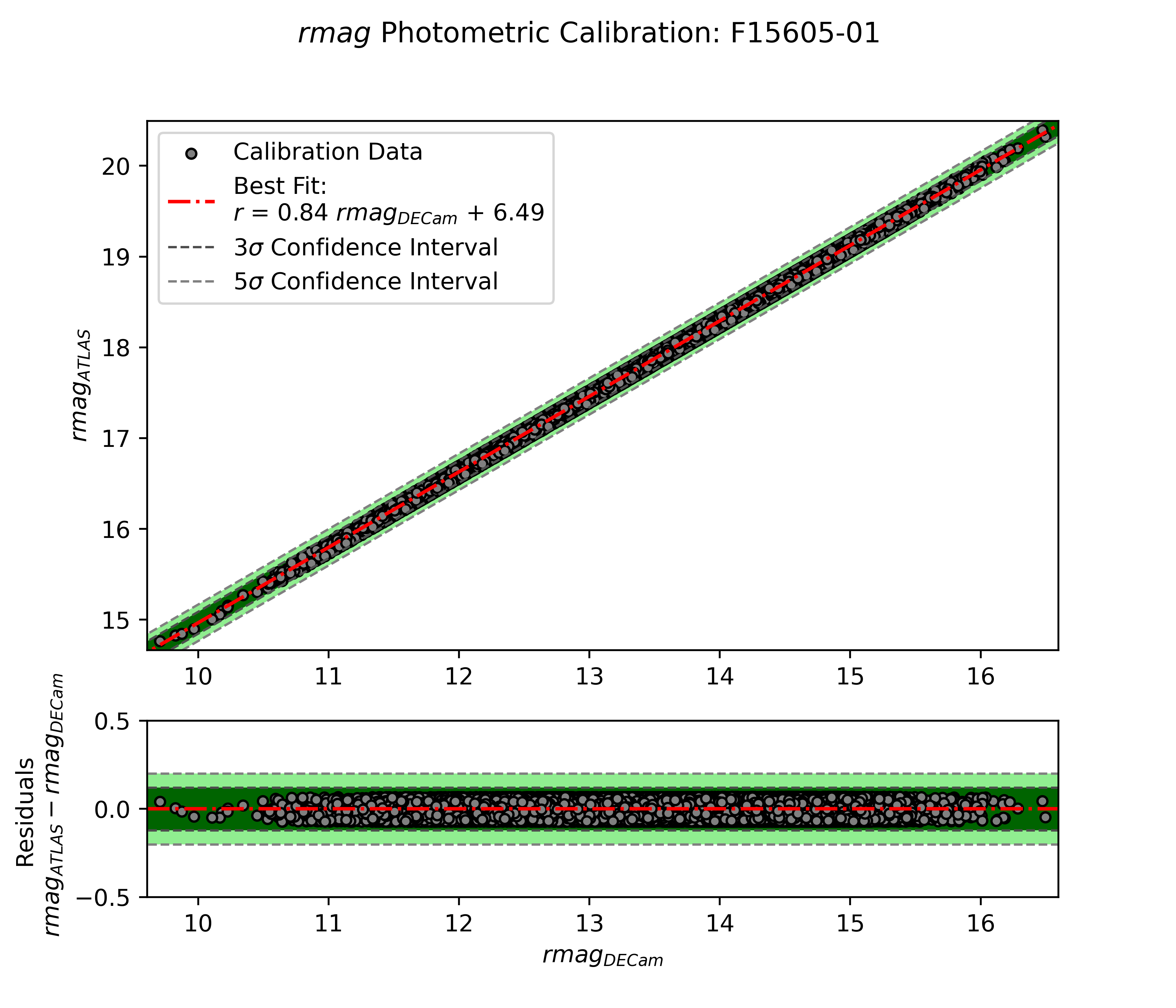}
    \caption{As in Figure~\ref{Fig::calib_g} but for the $r$-band images.}%
    \label{Fig::calib_r}%
\end{figure*}

\begin{figure*}[ht!]
    \centering
    \includegraphics[width=0.3\textwidth]{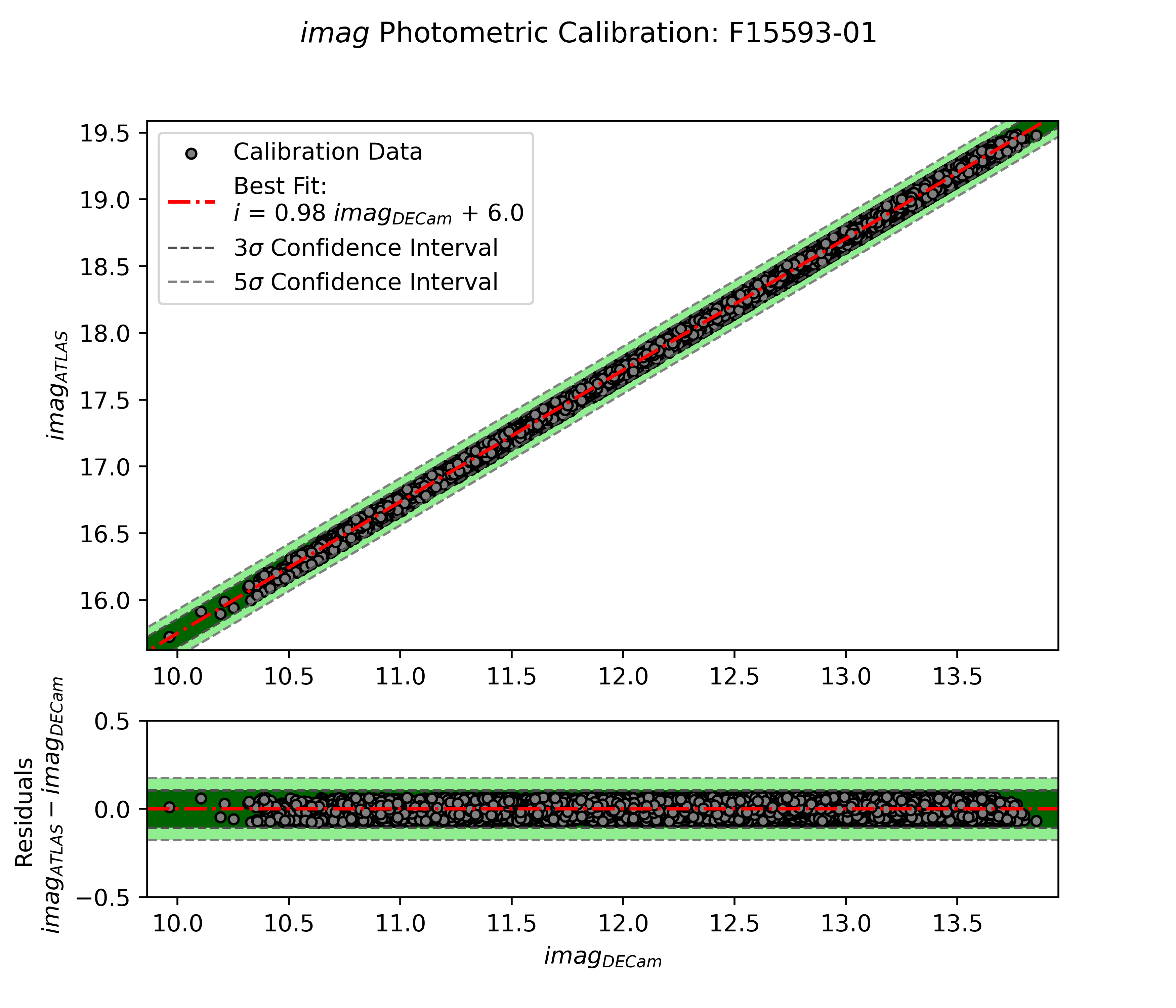}
    \includegraphics[width=0.3\textwidth]{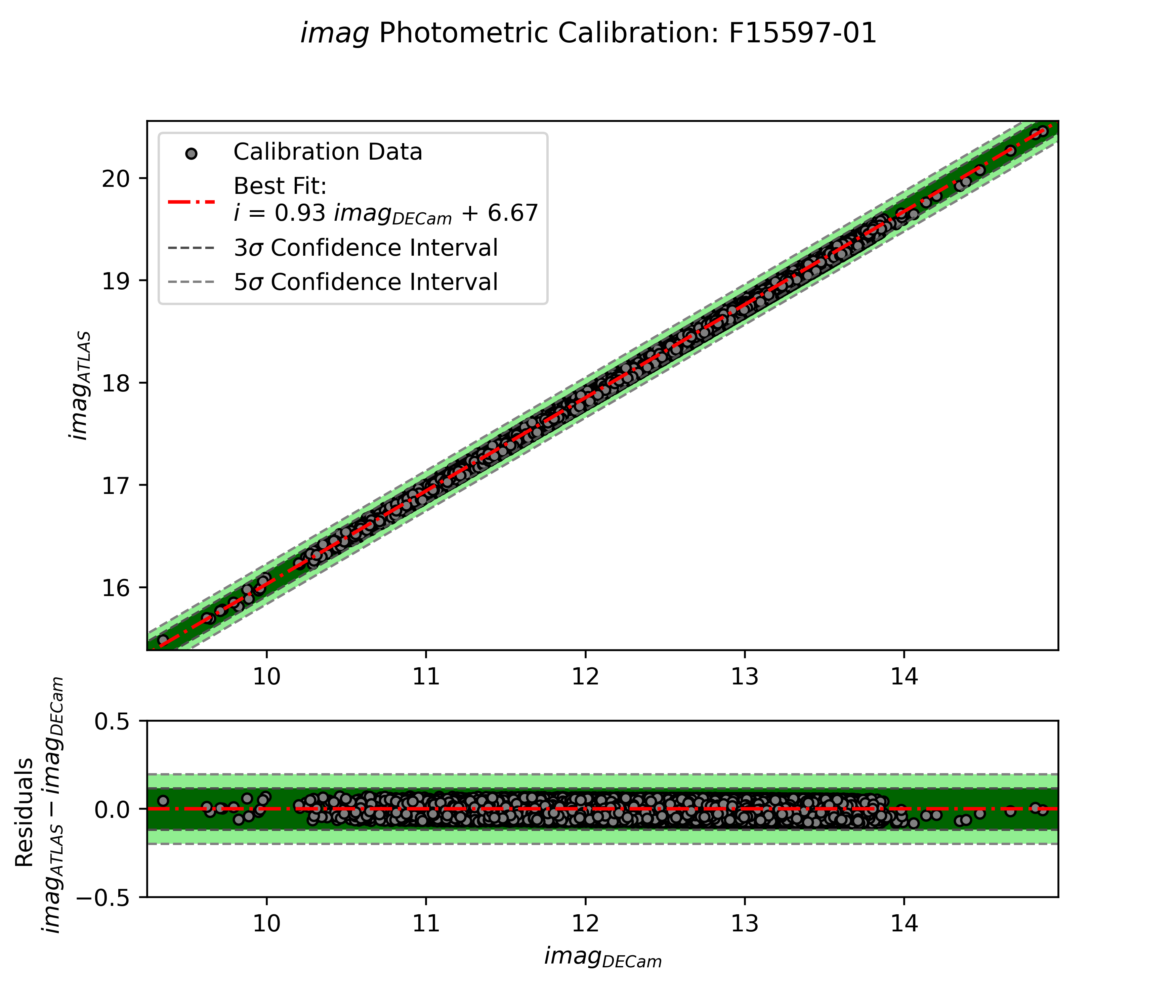}
    \includegraphics[width=0.3\textwidth]{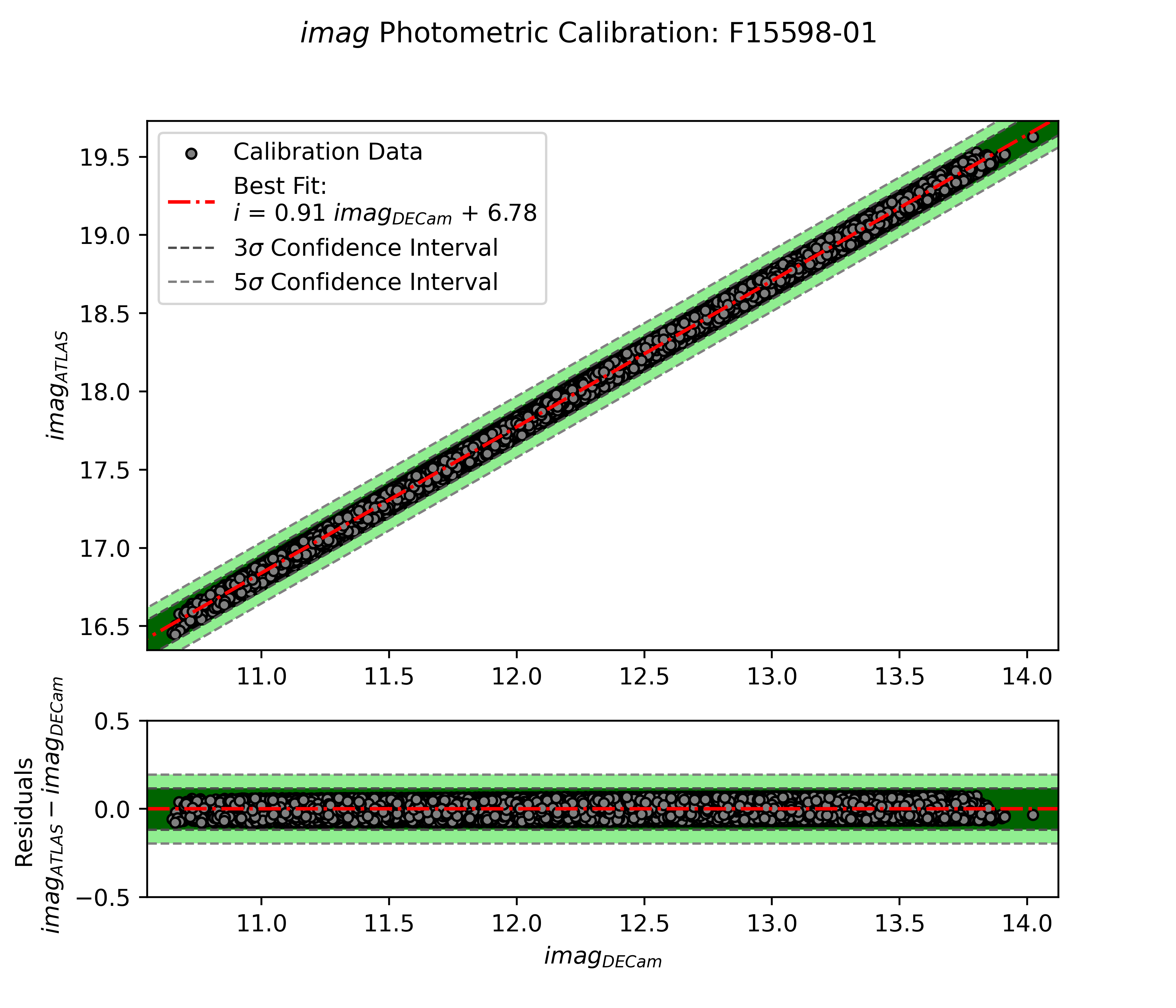}
    \includegraphics[width=0.3\textwidth]{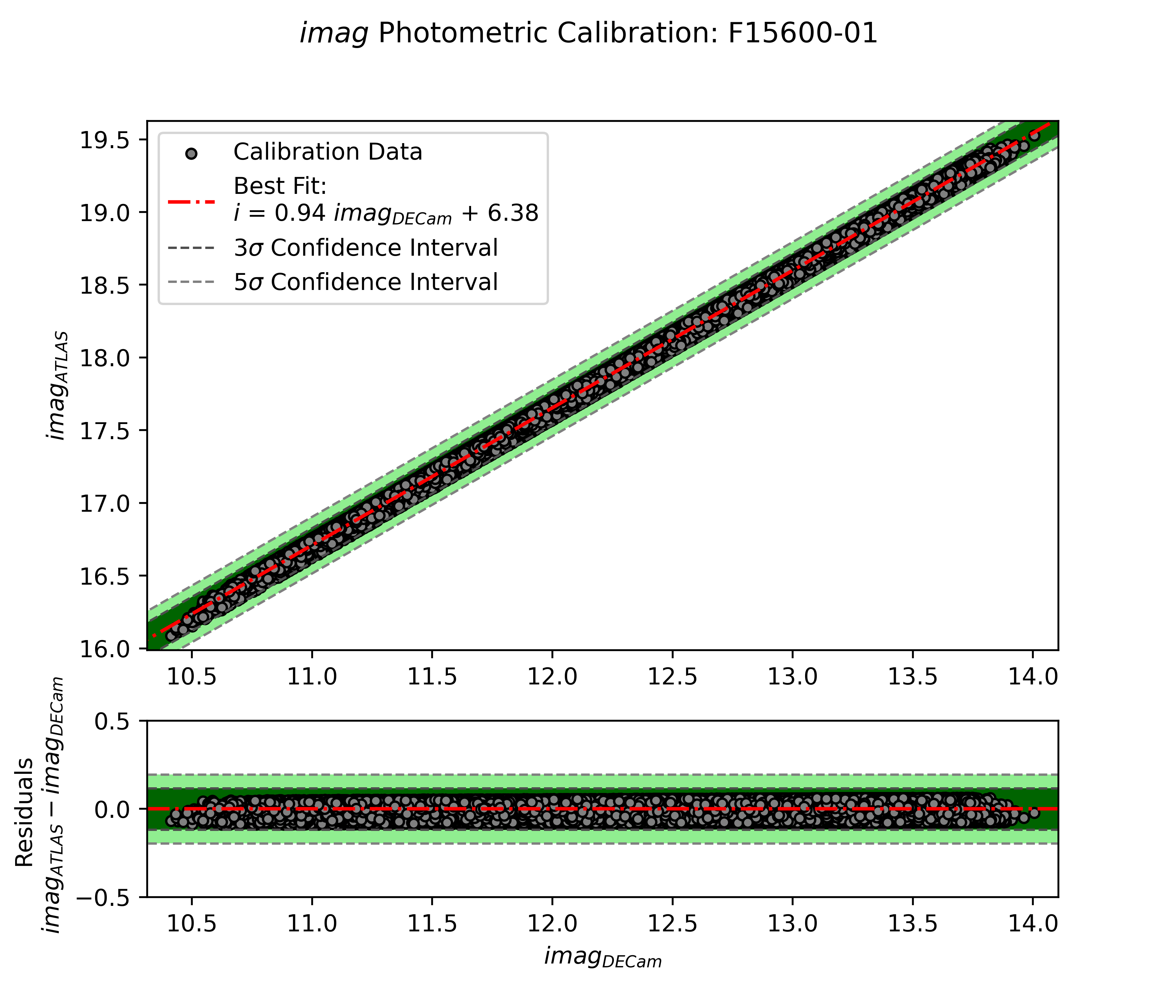}
    \includegraphics[width=0.3\textwidth]{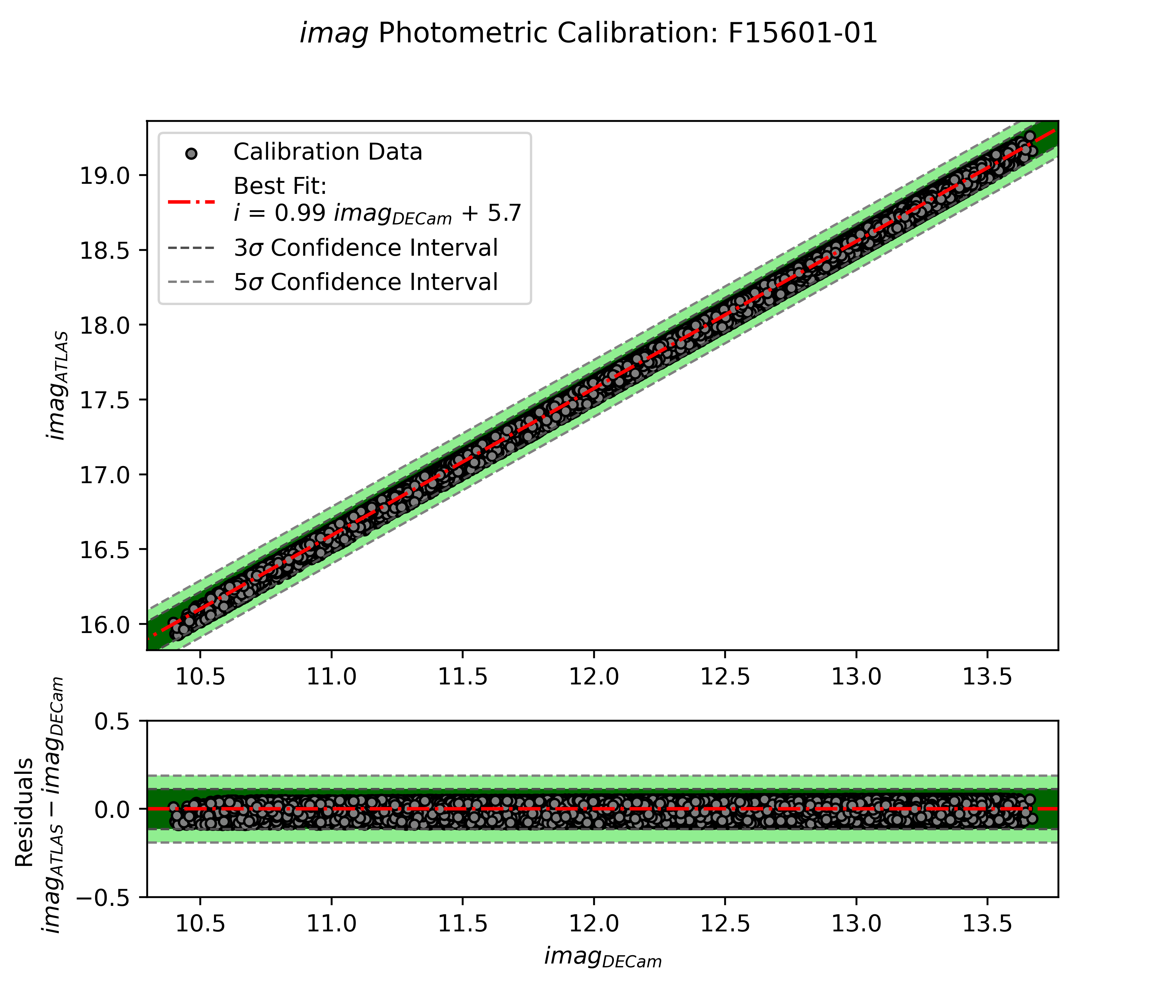}
    \includegraphics[width=0.3\textwidth]{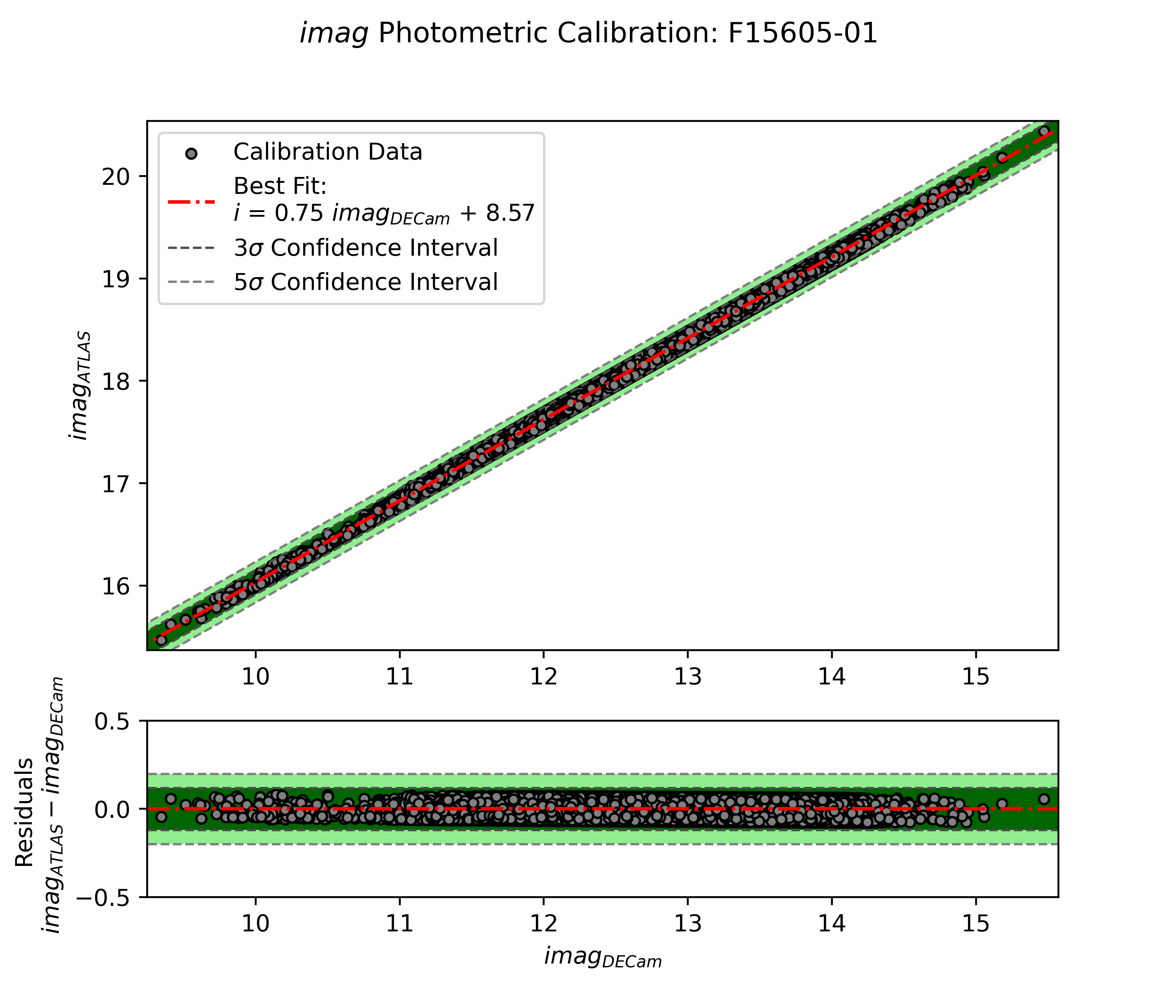}
    \caption{As in Figure~\ref{Fig::calib_g} but for the $i$-band images.
    \label{Fig::calib_i}}
\end{figure*}

The DES has been carried out by using the Dark Energy Camera \citep{flaugher2015}, designed and constructed by the DES Collaboration, which is a wide-field camera installed on the 4-meter V. Blanco Telescope at the Cerro Tololo Inter-American Observatory (CTIO) in Chile. DECam features a mosaic of 62 CCDs, providing a total resolution of 570 megapixels and a field of view of approximately 3.0~deg$^2$ per exposure. 

In this paper, we present the {\it photometric Catalogue of Observed Stars in the Small MagellanIc Cloud} (COSMIC-S in the following), including more than 10 million sources in the SMC region, with a completeness magnitude estimated to $m\sim22$ for all the investigated SDSS bands. The catalogue data are based on our analysis of the DECam images acquired between 2018 and 2020 in the $gri$ SDSS bands. In particular, in Section \ref{Sec::DataReduction} we describe the DECam instrument, the observations, the data reduction, and the photometric calibration in the SDSS system \citep{lenz1998}. In Section \ref{Sec::catalogue} we introduce the COSMIC-S catalogue, and present some useful plots and tables. Lastly, in Section \ref{Sec::conclusion} we give our final conclusions. 

\section{\label{Sec::DataReduction} Observations and data reduction}

Observations were performed in the $g$, $r$, and $i$ SDSS filters with DECam, mounted at the 4m V. Blanco Telescope at the CTIO (Chile). DECam is a 62 scientific CCDs module, 2000$\times$4000 pixels each, with a resolution of 0.27''/pixel, covering a sky area of roughly 3.0 deg$^2$ \citep{honscheid2008}.
Data used for the analysis carried out have been acquired during the 2018A-0273 DECam program (from February 2018 to January 2020) and are publicly available on the NOIRLab Archive website (see the \url{https://astroarchive.noirlab.edu/portal/search/} webpage). In particular, we used observations of the SMC region divided into six different DECam fields, each observed in the $gri$ passbands. In this way, we obtained full coverage of the SMC, with some gaps that have been considered and removed. The exposure times are 200, 100, and 200 seconds per band, respectively. 

The analysis has been performed using bias, dark and flat-field corrected images. Images have been analysed by our homemade pipeline, written in Python. We also used external software for the source extraction and the PSF extraction, i.e. SExtractor \citep{sextractor} and PSFEx \citep{psfex}. The implementation of a dedicated PSF extraction method is mandatory for such regions due to their high stellar density, excluding a priori diverse and minor precise techniques such as aperture photometry. The reduction pipeline consists of three main parts:

\subsection{PSF and source extraction}
    
First of all, we extracted bright and isolated sources by running SExtractor on DECam images, in order to produce a source catalogue suitable for the PSF extraction. In this specific case, we run SExtractor selecting its parameters to {\tt DETECT\_MINAREA=13}, {\tt DETECT\_THRESH=20} and {\tt ANALYSIS\_THRESH=20}, enforcing the detection of brighter objects only. We filtered the result by saving only {\tt FLAGS=0} stars, preventing spurious detection. We also checked that stars were not close to other sources, even fainter, or to the image borders by considering a minimum distance of 50 pixels. 

We then estimated the PSF for each image by running PSFEx and, consequently, we extracted again the brighter stars using SExtractor, by applying the same selection criteria as in the first run, but also considering the estimated PSF for a more accurate photometry. However, in this case, we did not apply any selection in distance because the current result is now used only for the photometric correction.

\subsection{Photometric Calibration}

The catalogue of bright sources obtained at the end of the previous stage contains instrumental magnitudes, in the $gri$ bands, that require a photometric calibration. For this purpose, we used The ATLAS All-Sky Stellar Reference Catalog \citep{tonry2018}, whose authors estimate its completeness to $m < 19$, which contains SDSS bands magnitudes for many stars, including those in the SMC region. Instrumental PSF magnitudes provided by SExtractor+PSFEx are then corrected by performing a robust linear regression, by using the RANSAC algorithm developed by \cite{ransac} and implemented in Python, and considering the photometric information given by \cite{tonry2018}. In order to have an even more solid photometric calibration, only matched sources with a magnitude difference scattering within one $\sigma$ are considered. This constraint gives the possibility to extract an accurate calibration relationship, for each DECam field and photometric band analysed. The performed photometric calibration is shown in Figs. \ref{Fig::calib_g},  \ref{Fig::calib_r}, and \ref{Fig::calib_i} for the $g$, $r$, and $i$ band, respectively. In the three figures, each panel contains two sub-boxes: in the upper one the DECam magnitudes (x-axis) and the corresponding ATLAS magnitudes (y-axis) are presented, while the bottom one shows the fit residuals. In both boxes, the two shades of green refer to the 3$\sigma$ (darker) and $5\sigma$ (lighter) confidence interval, while the red dash-dotted line is the best fit (corresponding to the zero residual line in the bottom box).

\subsection{PSF Photometry}

The last part of the procedure is devoted to the final PSF photometry considering both the PSF extracted by PSFEx and the photometric calibration mentioned in the previous item, applying both of them to all the sources detected by SExtractor. In this case, we run SExtractor by setting the crucial parameters {\tt DETECT\_MINAREA=3}, {\tt DETECT\_THRESH=1.5} and {\tt ANALYSIS\_THRESH=1.5}. Stars with bad PSF fitting, close to the CCD edges, or saturated sources with irregular profiles are discarded purging again the result by considering only sources with {\tt FLAGS=0}. At the end of this stage, the final catalogue is ready.

\begin{table*}
    \centering
    \begin{tabular}{llll} 
        \hline \hline
        \textbf{Column} & \textbf{Column} & \textbf{Units} & \textbf{Description}  \\
        \textbf{Number} & \textbf{Name} & &  \\
        \hline
        1 & COSMIC-S\_ID &  & Object Identification Number \\
        2 & RAJ2000 & deg & Right Ascension associated to the selected source \\
        3 & DEJ2000 & deg & Declination associated to the selected source \\
        4 & e\_RAJ2000 & deg & Uncertainty associated to the Right Ascension \\
        5 & e\_DEJ2000 & deg & Uncertainty associated to the Declination \\
        6 & gmag &   & SDSS g-band magnitude \\
        7 & e\_gmag &   & Uncertainty associated to the SDSS g-band magnitude \\
        8 & rmag &   & SDSS r-band magnitude \\
        9 & e\_rmag &   & Uncertainty associated to the SDSS r-band magnitude \\
        10 & imag &   & SDSS i-band magnitude \\
        11 & e\_imag &   & Uncertainty associated to the SDSS i-band magnitude \\
        \hline \hline
    \end{tabular}
    \caption{Columns description for the SMC photometric catalogue presented in this work. The first column represents the column name used in the catalogue, the second column specifies the physical units used for each information given in the catalogue, and the third column gives a short description of each catalogue column.}
    \label{Table::cat_info}
\end{table*}

\begin{figure}[!htp]
    \centering
    \includegraphics[width=\columnwidth]{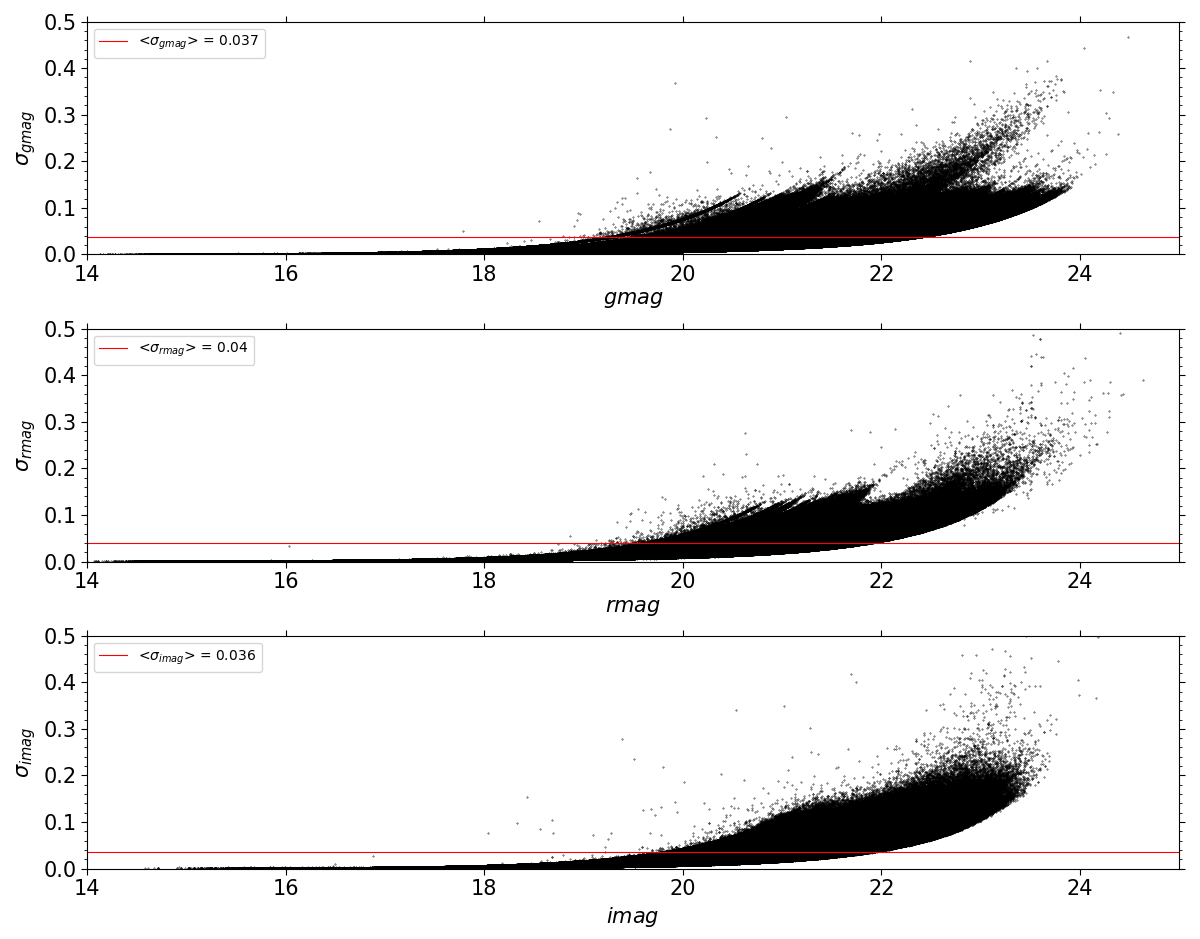}%
    \caption{Photometric errors versus magnitudes in the $g$, $r$, and $i$ bands (from the top panel to the bottom one) for all the stars in the COSMIC-S catalogue. The horizontal line in red marks the mean error.}
    \label{Fig::mag_vs_err}%
\end{figure}


\section{\label{Sec::catalogue} The COSMIC-S photometric catalogue}

The COSMIC-S final catalogue consists of 10 971 906 entries; of these, 2 456 434 have photometric information in all the $gri$ bands analysed. For each object, it provides an identification number (Column 1), celestial coordinates, i.e. Right Ascension and Declination and the corresponding uncertainties in the J2000 standard (columns 2-5), and the magnitudes in the $g$, $r$ and $i$ bands with associated the corresponding errors (columns 6-11). In Table \ref{Table::cat_info} we show the column names for each product given in the catalogue, specifying the physical units adopted and a short description.

Coordinate errors are given by SExtractor and are reported in the COSMIC-S catalogue as such.  Coordinates and errors are both expressed in degrees. The source magnitudes are obtained by performing a photometric calibration matching the DECam input catalogue, consisting of instrumental magnitudes as estimated by SExtractor and PSFEx, for each SDSS band, and the ATLAS catalogue. The output results in a list of corrected magnitudes with a given uncertainty corresponding to the statistical propagation between the SExtractor and the ATLAS errors, i.e. given by

\begin{equation}
    \sigma_{m} = \sqrt{(\sigma_m^{SEx})^2 + (\sigma_m^{ATLAS})^2}
\end{equation}

\begin{figure}[!htp]
    \centering
    \includegraphics[width=\columnwidth]{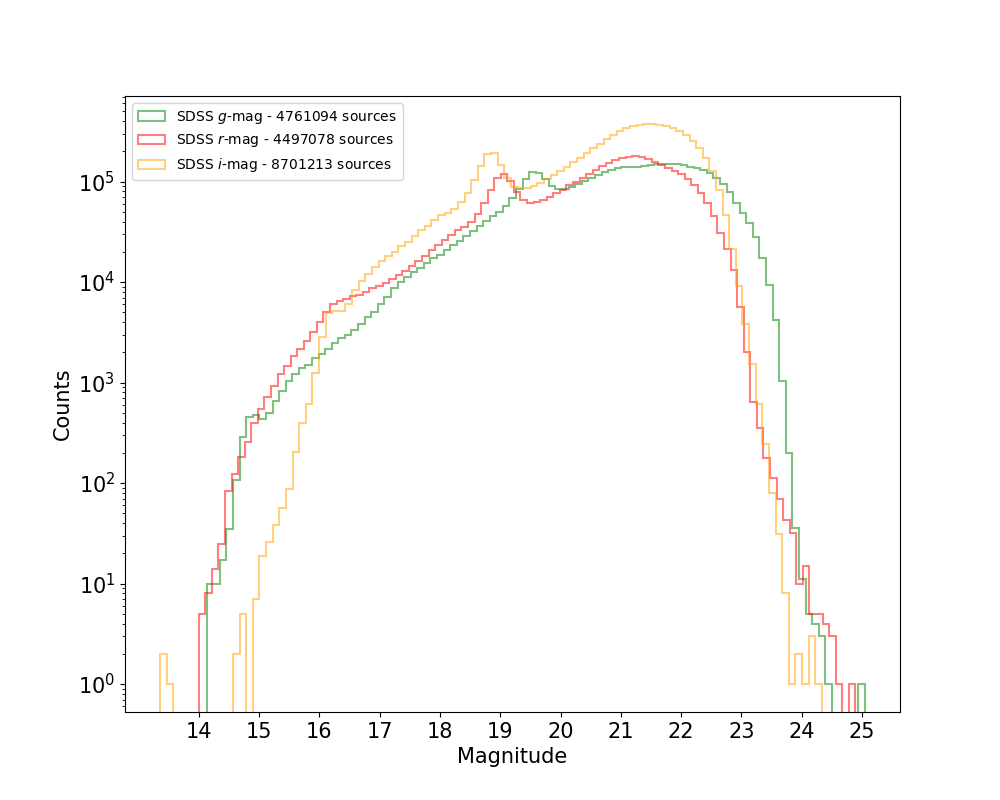}%
    \caption{Magnitude distributions for the stars detected in each photometric band considered in this work, i.e. $g$ in green, $r$ in red and $i$ in yellow. In the top left box, the legend is reported, indicating the number of detected sources for each band. Completeness and limiting magnitude derived from these histograms are reported, for simplicity, in Table \ref{Table::compl_lim}.}
    \label{Fig::mag_hist}%
\end{figure}

Figure~\ref{Fig::mag_vs_err} shows the photometric errors versus the magnitude in the three bands considered. In the three panels, the red horizontal line represents the mean value of the corresponding error, resulting in $\sigma_{gmag}=0.037$, $\sigma_{rmag}=0.04$, and $\sigma_{imag}=0.036$. The magnitudes distributions for the $g$ (green), $r$ (red) and $i$ (yellow) filters are shown in Figure~\ref{Fig::mag_hist}. The histogram peak, located before the rough drop in the magnitudes distribution, gives an estimate of the completeness magnitude while the limiting magnitudes are estimated by considering the magnitude x-range. In particular, these values are critical for a good photometric catalogue, especially for discriminating stars at the distance of the SMC, i.e. $\sim60$~kpc, with respect to background or foreground objects. The extracted values are listed in Table \ref{Table::compl_lim}. In the case of COSMIC-S, the catalogue results complete for magnitudes $m_c \lesssim 21.5$, with a detection limit $m_l\sim 25$ for all three bands, which is a remarkable result considering the ground-based nature of the telescope used for the observations.

\begin{table}[!htp]
    \centering
    \begin{tabular}{ll} 
        \hline \hline
        \textbf{Data product} & \textbf{Number of sources} \\
        \hline
        Total sources & 10 971 906 \\
        $g$-band magnitude &  4 761 094 \\
        $r$-band magnitude &  4 497 078 \\
        $i$-band magnitude &  8 701 213 \\
        $gri$ mag & 2 456 434 \\
        \hline \hline
    \end{tabular}
    \caption{Number of total, single-band, and $gri$-bands sources sources available in the COSMIC-S released catalogue.}
    \label{Table::number_of_products}
\end{table}

\begin{table}[!htp]
    \centering
    \begin{tabular}{lll} 
        \hline \hline
        \textbf{Magnitude} & \textbf{Completeness} & \textbf{Limiting} \\
         \textbf{band} & \textbf{magnitude} & \textbf{magnitude} \\
        \hline
        $g$-mag & 21.55 & 25.05 \\
        $r$-mag & 21.18 & 24.89 \\
        $i$-mag & 21.48 & 24.33 \\
        \hline \hline
    \end{tabular}
    \caption{Completeness and limiting magnitudes for the SMC stars in the $gri$ photometric bands considered in this work. The completeness value is extracted as the magnitude corresponding to the peak from the histogram in Figure~\ref{Fig::mag_hist}. The limiting magnitude is estimated as the highest magnitude associated with a source detection.}
    \label{Table::compl_lim}
\end{table}

\begin{figure*}[ht!]
    \centering
    \includegraphics[width=0.49\textwidth]{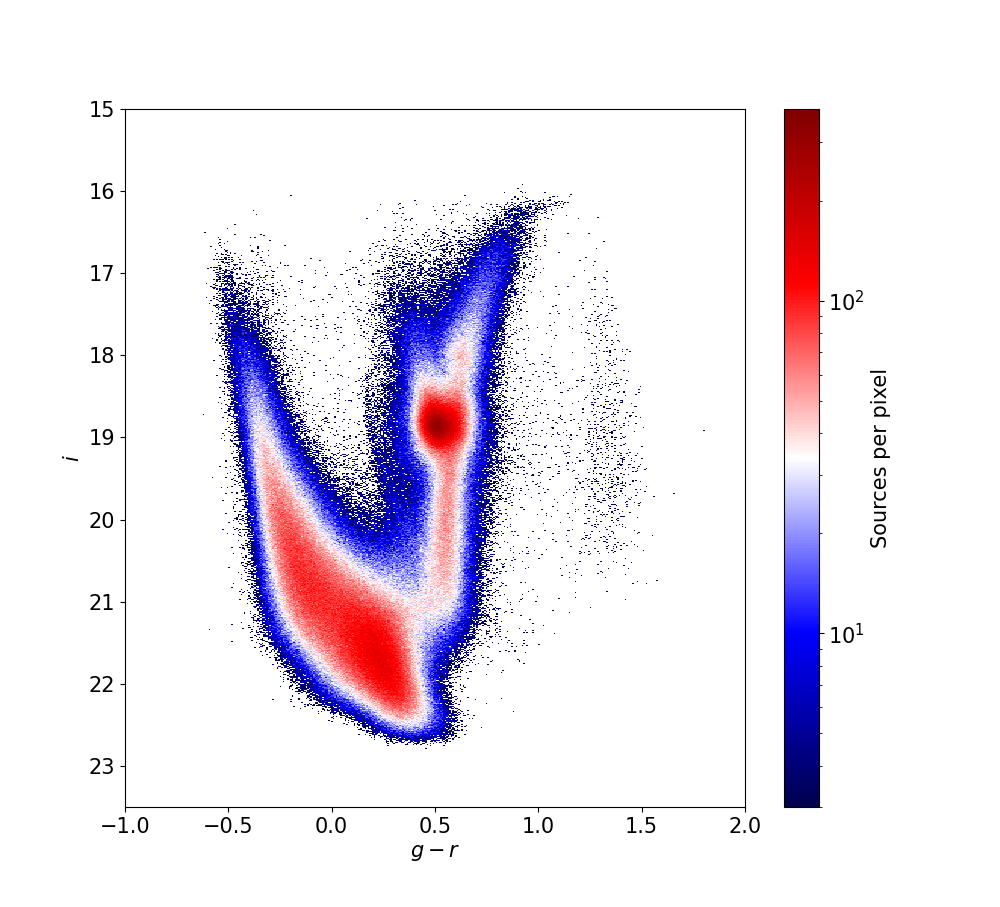}
    \includegraphics[width=0.49\textwidth]{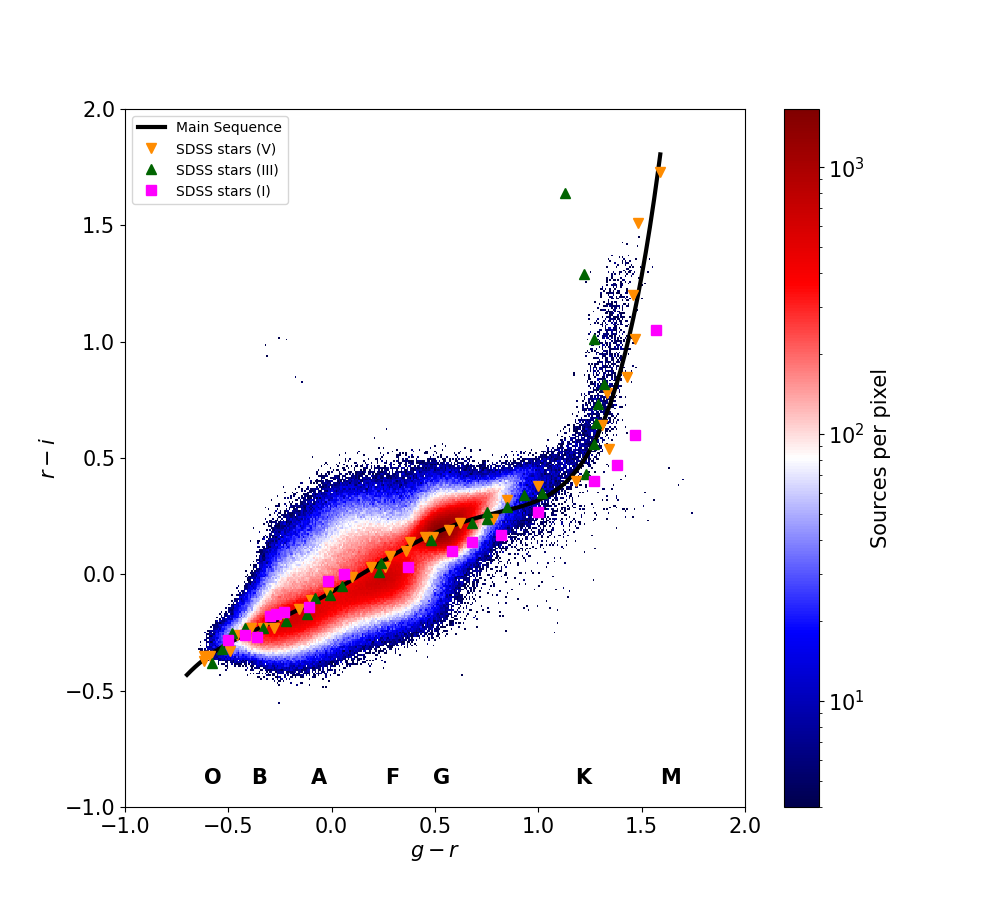}
    \caption{Left panel: colour-magnitude ($i$ vs $g-r$) diagram for the SMC sources. The main sequence (left) and the giant branch (right) are clearly distinguishable. The clustered region for $g-r>1.2$ shows foreground M-type and K-type stars, located in the Milky Way.
    Right panel: colour-colour ($r-i$ vs $g-r$) diagram for the SMC sources. The mean spectral class at different colours is indicated by the conventional O, B, A, F, G, K, M letters at the bottom of the figure. Luminosity classes associated to the reported spectral class are plotted in orange (V), green (III), and fuchsia (I). The black solid line follows the main sequence. The total number of sources plotted is 2~456~5434, as also indicated in Table \ref{Table::number_of_products}. The 2D histograms are built considering 1000 bins and discarding all groups with a number of sources per pixel less than 3, in order to avoid the display of excessive noise. The colorbar shown in the right of each panel gives the sources/pixel density of the plotted pixels.}%
    \label{Fig::hr_diag}%
\end{figure*}

\begin{figure}[ht!]
    \centering
    \includegraphics[width=\columnwidth]{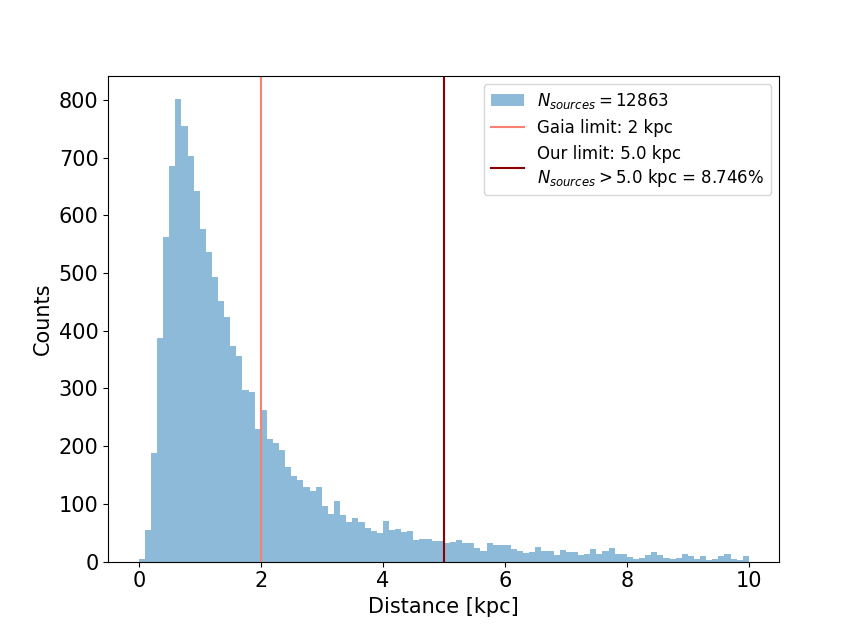}%
    \caption{Distance distribution for the 12~863 stars with $g-r>1.2$ from the Figure~\ref{Fig::hr_diag} obtained from the Gaia EDR3 catalogue, with distance measurements corrected by \cite{bailerjones2021}. The Gaia detection limit is indicated by the pink vertical line while our extended limit, as used in this work, is reported with the red vertical line. In the top-right panel the legend is also illustrated. Most of the sources ($\simeq 90\%$) are within 5~kpc from Earth, representing then a group of local stars, located in both the thin and thick disks.}
    \label{Fig::dist_hist}%
\end{figure}

The colour-magnitude and colour-colour diagrams in Figure~\ref{Fig::hr_diag} clearly show the main sequence and giant branch clusters. Colour-magnitude ($i$ vs $g-r$) diagram for the SMC sources in the left panel of Figure~\ref{Fig::hr_diag} shows the main sequence branch (left) and the giant branch (right). The densest and reddest region of the diagram located at $g-r\simeq 0.5$ and $i\simeq 19$ corresponds to the Red Clump stars in the SMC. In the left panel of Figure \ref{Fig::hr_diag} it is also visible a cluster of objects with $1.2 \lesssim g-r \lesssim 1.6$. It is composed by 12~863 stars and our analysis shows that they are most likely foreground K-type and M-type stars. Indeed, we recovered their distance by considering the Gaia~EDR3 catalogue and the corresponding estimated distances provided by \cite{bailerjones2021}. In Figure~\ref{Fig::dist_hist} we give the distance distribution for these selected stars. As shown in the figure only $\lesssim 9\%$ of these  sources have distances greater than about 5~kpc (red vertical line). In particular, the Gaia collaboration in the DR3 \citep{gaiadr3} quotes the distance limit in $\simeq 2$~kpc (see, e.g., \citealt{andrae2023}), shown as a pink vertical line in Figure~\ref{Fig::dist_hist}, beyond which the discrepancy between the distance evaluated by Gaia and the corresponding one in the literature becomes gradually larger. In this work we consider a less conservative limit, putting this on a threshold of $\simeq 5$~kpc. 
This introduced limit consists on the fact that, when observing in the direction of the SMC, we include many foreground objects, due to the inclination of the SMC with respect to the galactic plane, so that the galactic disk crossed, consisting of both the thin and thick one, results of $\simeq5$~kpc.  We than conclude that objects lying in this region of the colour-magnitude diagram are Milky Way's stars placed either in the thin and thick disk. 
In the colour-colour ($r-i$ vs $g-r$) diagram in the right of Figure~\ref{Fig::hr_diag} the Red Clumps region is located at $g-r\simeq0.5$ and $r-i\simeq0.2$. The rapid change of inclination for $g-r\gtrsim1.2$ shows the position of K-type and M-type stars in the diagram, correlated to the isolated cluster of stars in the colour-magnitude diagram (left panel in Figure~\ref{Fig::hr_diag}). The plot also reports the spectral (black bold letters in the bottom) and luminosity classes (V in orange, III in green, and I in fuchsia) taken from \cite{covey2007}, also indicating the main sequence line in black.


\section{\label{Sec::conclusion} Conclusions}

In this work, we presented the COSMIC-S photometric catalogue, the result of an intense analysis of the SMC images acquired by the Dark Energy Camera by using the $gri$ SDSS filters. We developed a pipeline for the photometric calibration and analysis producing the COSMIC-S catalogue, containing more than 10 million objects with astrometric and photometric features. We produced the colour-magnitude and colour-colour diagrams in order to test the effective goodness of the extracted photometry and to compare this result to what is present in the literature. We identified a cluster of stars in the colour-magnitude diagram that are most likely K-type and M-type foreground stars, thanks to the distances provided by the Gaia EDR3 catalogue. As discussed in detail in the manuscript, the COSMIC-S catalogue appears virtually complete to magnitudes $m\simeq22$ in the three $gri$ photometric bands, reaching a detection limit $\simeq 25$. Moreover, it is worth noting that many photometric catalogues in the literature, including the ATLAS catalogue used in this work as the reference one for the performed photometric calibration, have a lower completeness magnitude, which is $m\simeq 19$ for the ATLAS catalogue in the three SDSS bands. This result makes our product one of the deeper and most complete photometric catalogues of the SMC field of view, compared to what the scientific literature offers. 


\section*{Acknowledgements}

This paper is based on publicly available observations by DECam (Dark Energy Camera), an instrument mounted on the V. Blanco Telescope, as part of the Cerro Tololo Inter-American Observatory (Chile). DECam images used for this work are publicly available at the \url{https://astroarchive.noirlab.edu/portal/search/} webpage.
\noindent We thank for partial support the INFN projects TAsP and Euclid.


\end{document}